\documentclass[prl,reprint]{revtex4-1}
\pdfoutput=1
\usepackage{graphicx}
\usepackage{color}
\usepackage{bm}
\usepackage{hyperref}

\begin{document}

\title{Fractional Quantum Hall Effect at $\nu=1/2$ in Hole Systems Confined to GaAs Quantum Wells}
\date{today}

\author{Yang Liu}
\affiliation{Department of Electrical Engineering,
Princeton University, Princeton, New Jersey 08544}
\author{A.L.\ Graninger}
\affiliation{Department of Electrical Engineering,
Princeton University, Princeton, New Jersey 08544}
\author{S.\ Hasdemir}
\affiliation{Department of Electrical Engineering,
Princeton University, Princeton, New Jersey 08544}
\author{M.\ Shayegan}
\affiliation{Department of Electrical Engineering,
Princeton University, Princeton, New Jersey 08544}
\author{L.N.\ Pfeiffer}
\affiliation{Department of Electrical Engineering,
Princeton University, Princeton, New Jersey 08544}
\author{K.W.\ West}
\affiliation{Department of Electrical Engineering,
Princeton University, Princeton, New Jersey 08544}
\author{K.W.\ Baldwin}
\affiliation{Department of Electrical Engineering,
Princeton University, Princeton, New Jersey 08544}
\author{R. Winkler}
\affiliation{Department of Physics,
Northern Illinois University, DeKalb, Illinois 60115}
\affiliation{Materials Science Division, Argonne National Laboratory,
  Argonne, Illinois 60439, USA}

\date{\today}

\begin{abstract}
  We observe fractional quantum Hall effect (FQHE) at the
  even-denominator Landau level filling factor $\nu=1/2$ in
  two-dimensional hole systems confined to GaAs quantum wells of width
  30 to 50 nm and having bilayer-like charge distributions. The
  $\nu=1/2$ FQHE is stable when the charge distribution is symmetric
  and only in a range of intermediate densities, qualitatively similar
  to what is seen in two-dimensional electron systems confined to
  approximately twice wider GaAs quantum wells. Despite the complexity
  of the hole Landau level structure, originating from the coexistence
  and mixing of the heavy- and light-hole states, we find the hole
  $\nu=1/2$ FQHE to be consistent with a two-component,
  Halperin-Laughlin ($\Psi_{331}$) state.
\end{abstract}


\maketitle

In a large perpendicular magnetic field ($B$), interacting
two-dimensional electron systems (2DESs) exhibit fractional quantum
Hall effect (FQHE) \cite{Tsui.PRL.1982}, predominantly at
odd-denominator Landau level (LL) filling factors $\nu$
\cite{Jain.CF.2007}. Much more rarely, FQHE is observed at
\emph{even-denominator} fillings such as $\nu=5/2$
\cite{Willett.PRL.1987} and 1/2 \cite{Suen.PRL.1992,
  Eisenstein.PRL.1992}. The origin of these states has long been
enigmatic. The 5/2 FQHE is likely to be the one-component, Moore-Read
(Pfaffian) state \cite{Moore.Nuc.Phy.1991}, a state that is expected
to obey non-Abelian statistics and find use in topological quantum
computing \cite{Nayak.Rev.Mod.Phys.2008}. Alternatively, it might be a
two-component, Abelian, Halperin-Laughlin ($\Psi_{331}$) state
\cite{Halperin.HPA.1983}, if the spin degree of freedom is
necessary. The $\nu=1/2$ FQHE has been reported in 2DESs confined
either to double quantum wells (QWs) \cite{Eisenstein.PRL.1992}, or to
symmetric, wide, single QWs where the electrons typically occupy two
electric subbands and have a bilayer-like charge distribution
\cite{Suen.PRL.1992}. It is generally believed to be a $\Psi_{331}$
state, with the two pseudo-spin components being the symmetric and
antisymmetric electric subbands or, equivalently, the two "layers"
which can be constructed form a linear combination of the
subbands. The $\Psi_{331}$ state is stable when the Coulomb energy is
much larger than the pseudo-spin splitting (the subband energy
separation), and the inter-layer and intra-layer Coulomb energies are
comparable \cite{Yoshioka.PRB.1989, Eisenstein.PRL.1992,
  Suen.PRL.1992, Suen.PRL.1992b, He.PRB.1993, Suen.PRL.1994,
  Papic.PRB.2009a, Papic.PRB.2009, Peterson.PRB.2010,
  Peterson.PRB.2010b, Shabani.PRB.2013}. While these conditions are
clearly met for the $\nu=1/2$ FQHE observed in double QWs, the case
for the wide QWs is less clear: The subband separation in wide QWs is
large and in principle a one-component, Pfaffian state at $\nu=1/2$
can be stablized \cite{Greiter.PRL.1991}.

Up to now, the $\nu=1/2$ FQHE has only been observed in GaAs 2DESs
and, very recently, in bilayer graphene
\cite{Ki.Morpurgo.cond.mat.2013}. Here we report the first observation
of FQHE at $\nu=1/2$ in very high-quality 2D \emph{hole} systems
(2DHSs) confined to relatively wide GaAs QWs. Thanks to the
coexistence of heavy-hole (HH) and light-hole (LH) states, as well as
the strong spin-orbit interaction and HH-LH mixing, the LL structure
of 2DHSs confined to wide QWs is quite complex and non-linear and it
includes multiple level crossings and anti-crossings \cite{Note1}. Despite
these complexities, far from the LL crossings, the hole $\nu=1/2$ FQHE
exhibits qualitatively the same evolution as in 2DESs.  It is only
stable in symmetric QWs, emerges from a compressible state at low
densities, and is destroyed by an insulating phase at high
densities. Based on a detailed discussion of the data, we conclude
that, although we cannot rule out the possibility that the strong LL
mixing may stablize a single-component 1/2 state, the $\nu=1/2$ FQHE
in our 2DHSs is consistent with a two-component $\Psi_{331}$ state.

Our samples are made from GaAs wafers grown by molecular beam epitaxy
along the (001) direction. The 2DHSs are confined to GaAs QWs with
widths $W=$ 30 to 50 nm, flanked by undoped
Al$_{0.3}$Ga$_{0.7}$As spacer layers and carbon $\delta$-doped layers,
and have a very high mobility $\mu \gtrsim$ 200 m$^2$/Vs at low
temperatures. We made 4 $\times$ 4 mm$^2$ samples in a van der Pauw
geometry with alloyed In:Zn contacts at their four corners. We then
fitted each sample with an evaporated Ti/Au front-gate and an In
back-gate to control the charge distribution symmetry in the QW and 2D
hole density, $p$, which we give throughout this letter in units of
$10^{11}$ cm$^{-2}$. All the measurements were carried out in dilution
refrigerators at their base temperature ($T \simeq$ 30 mK).

\begin{figure*}
\includegraphics[width=.9\textwidth]{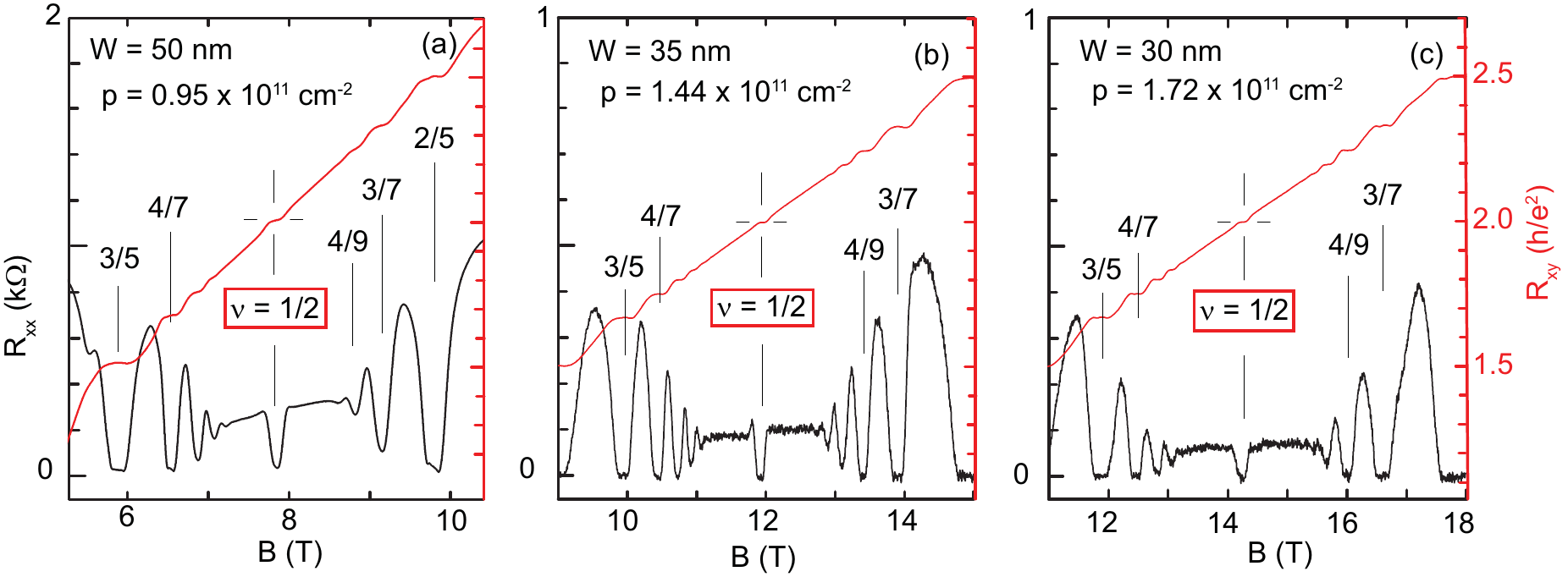}
\caption{\label{fig:cartoon}(color online) Longitudinal (${R_{xx}}$)
  and Hall ($R_{xy}$) resistance vs perpendicular magnetic field ($B$)
  traces taken at $T\simeq30$ mK for the 50-, 35- and 30-nm-wide QWs
  at densities $p = 0.95$, 1.44 and 1.72$\times 10^{11}$ cm${^{-2}}$,
  respectively.}
\end{figure*}

Figure 1 highlights our main finding: the observation of $\nu=1/2$
FQHE in 2DHSs confined to GaAs QWs. Data are shown for QWs with
different $W$ and $p$. The $\nu=1/2$ FQHE is evidenced by a very deep
minimum in the longitudinal resistance ($R_{xx}$) at $\nu=1/2$,
accompanied by a Hall resistance ($R_{xy}$) plateau, well-quantized at
$2h/e^2$.

\begin{figure}
\includegraphics[width=.45\textwidth]{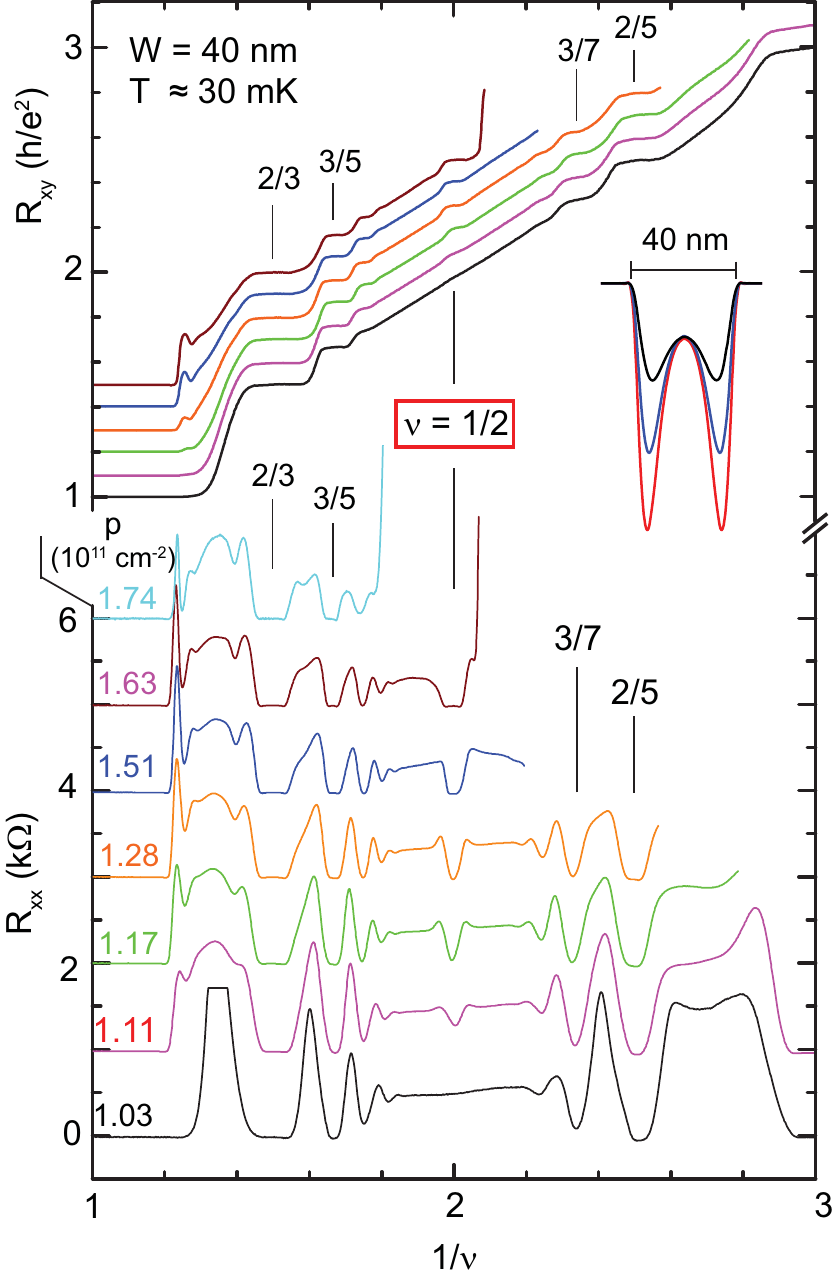}
\caption{\label{fig:waterfall} (color online) Waterfall plots of
  ${R_{xx}}$ and $R_{xy}$ vs $1/\nu$ for the 40-nm-wide QW as $p$ is
  changed from 1.03 to 1.74, while keeping the total hole charge
  distribution symmetric. Traces are shifted vertically for
  clarity. The $\nu=1/2$ FQHE is not seen at the lowest $p$, and is
  replaced by an insulating phase at the highest $p$. The inset shows
  the calculated charge distribution for $p=1.0$ (black), 1.5 (blue),
  and 2.0 (red).}
\end{figure}

Figure 2 shows a waterfall plot of $R_{xx}$ and $R_{xy}$ vs $1/\nu$
traces for $p=1.03$ to 1.74 taken in the 40-nm-wide QW. At the lowest
$p$ the 2DHS is compressible at $\nu=1/2$ and FQH states are seen only
at odd-denominator fillings $\nu=$ 1/3, 2/5, 3/7, ..., and 2/3, 3/5,
4/7, ..., similar to what is seen in high-quality, single-subband, 2D
electron or hole systems confined to narrow QWs. When we slightly
increase $p$ to $\simeq 1.11$, a FQHE emerges at $\nu=1/2$, indicated
by a dip in $R_{xx}$ and a developing $R_{xy}$ plateau. The $R_{xx}$
minimum gets deeper and the $R_{xy}$ plateau becomes flatter and wider
as we continue to raise the density, indicating an increasingly
stronger $\nu=1/2$ FQHE. At $p=1.51$, an insulating phase (IP) starts
to emerge at low fillings (high fields) and moves towards higher
fillings with increasing density. The IP reaches just to the
low-filling side of the $\nu=1/2$ FQHE at $p=1.63$, and then at
slightly higher density it destroys the $\nu=1/2$ FQHE as it keeps
moving to higher fillings.

\begin{figure}[htb]
\includegraphics[width=.4\textwidth]{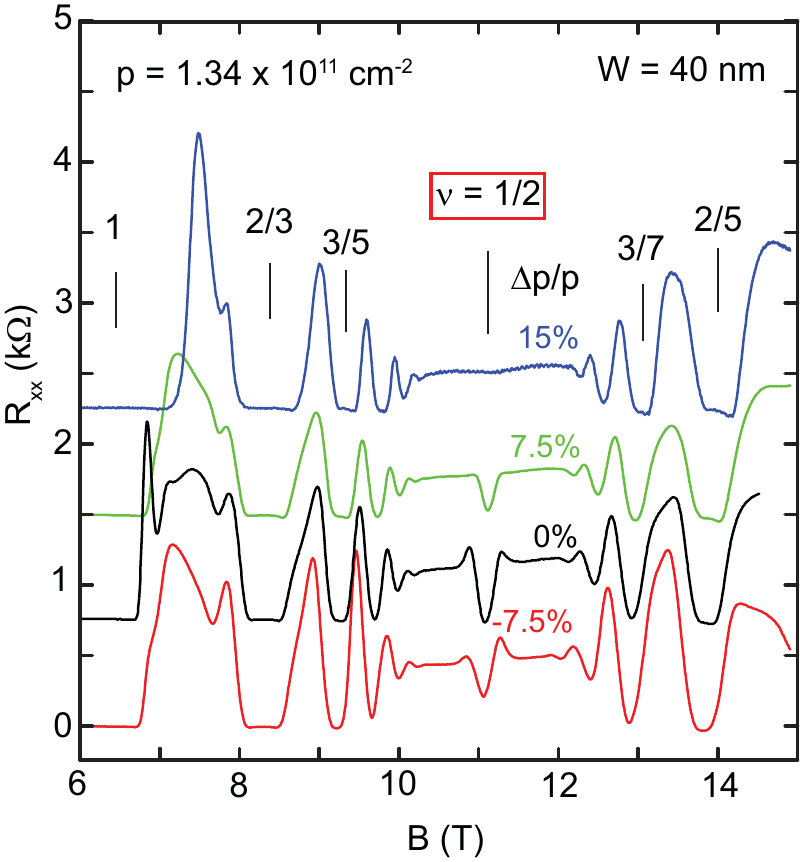}%
\caption{\label{fig:colorful} (color online) ${R_{xx}}$ vs $B$ for the
  40-nm-wide QW at a fixed density but different QW symmetries induced
  via applying front- and back-gate voltage biases.}
\end{figure}

The evolution seen in Fig. 2 is qualitatively very similar to the one
seen in 2DESs confined to wide GaAs QWs \cite{Suen.PRL.1994,
  Manoharan.PRL.1996}, with two notable exceptions in the FQHE to IP
transition at high densities. First, in 2DESs, at the highest
densities, when the system becomes very much bilayer-like, the IP
moves to fillings near $\nu=2/5$ \cite{Manoharan.PRL.1996}, namely 1/5
in each layer. This was attributed to the formation of a pinned Wigner
crystal (WC); note that 1/5 is indeed the filling near which a pinned
electron WC is observed in high-quality, single-layer 2DESs
\cite{Jiang.PRL.1990, Goldman.PRL.1990, Shayegan.PQHE.1998}. In the
2DHS, on the other hand, at high densities, the IP moves to fillings
larger than 2/5. In another 40-nm-wide QW sample in which we can reach
higher densities, the IP eventually approaches the $\nu=2/3$ FQHE at
$p\simeq 2.1$ when the system becomes quite bilayer-like. This is not
unexpected. Low-density, single-layer 2DHSs are in fact known to
exhibit an IP near $\nu=1/3$, interpreted as a pinned hole WC
\cite{Santos.PRL.1992, Engel.PRL.1997, Shayegan.PQHE.1998}. The larger
in-plane effective mass in GaAs 2DHSs, which causes significant LL
mixing and renders the 2DHS effectively more dilute, is believed to be
responsible for the higher $\nu$ near which the hole WC forms
\cite{Santos.PRL.1992, Shayegan.PQHE.1998}. Second, in 2DESs, for some
particular density the IP is seen on both sides of a strong $\nu=1/2$
FQHE \cite{Manoharan.PRL.1996}. In the 2DHS, however, we do not see
such a reentrant IP. At $p=1.63$ the IP is observed only on the
low-filling side of the $\nu=1/2$ FQHE. As $p$ is slightly increased
above 1.63, the $\nu=1/2$ FQHE is quickly destroyed and the IP takes
over at and near $\nu=1/2$ (see the $p=1.74$ trace in Fig. 2).

Returning to the $\nu=1/2$ FQHE, in Fig. 3 we study this state in the
40-nm-wide QW at fixed $p = 1.34$ with different QW charge
distribution symmetries. We change the symmetry by increasing the
density by $\Delta p$ via applying front-gate bias and decreasing the
density by $\Delta p$ via back-gate bias. The $\nu=1/2$ FQHE is strong
when the QW is symmetric ($\Delta p/p=0$). It becomes weak when
$\Delta p/p = \pm 7.5\%$ and is completely destroyed when $\Delta p/p
= 15\%$. This evolution is also very similar to what is observed in
2DESs confined to wide GaAs QWs. For example, in a 77-nm-wide QW, the
$\nu=1/2$ FQHE is strong at density $n=1.15$ when the QW is symmetric,
and disappears when $\Delta n/n \gtrsim 7\%$ \cite{Suen.PRL.1994}.

The data presented so far indicate that the $\nu=1/2$ FQHE in 2DHSs
exhibits qualitatively the same features as in 2DESs: It is only
stable in symmetric QWs and at intermediate densities. For a closer,
more quantitative comparison, in Fig. 4(a) we show a $W$ vs density
phase diagram for the ground state at $\nu=1/2$ in 2DESs confined to
wide GaAs QWs \cite{Shabani.PRB.2013}. The yellow band is the region
where the FQHE is the ground state of the 2DES. The 2DES becomes
compressible in the lower left (green) region, and exhibits an IP in
the upper right (red) region. In Fig. 4(a) we also show the 2DHS data
by red symbols, but after shifting the $y$-axis ($W$) upward by 34
nm. The solid symbols represent the parameters for which we observe a
$\nu=1/2$ FQHE, and the open symbols represent the compressible phase
(at low densities) or the IP (at high densities) in our 2DHSs. It it
clear that, if we include the 2DHS data in Fig. 4(a) diagram without
any adjustments in the QW width, all the data points would fall well
below the yellow band and deep in the compressible
region. Nevertheless, we will argue that the hole $\nu=1/2$ FQHE in
our 2DHSs has the same origin and is likely a two-component
$\Psi_{331}$ state.

First, the extreme sensitivity of the $\nu=1/2$ FQH state to the 2DHS
charge distribution symmetry (Fig. 3) provides clear evidence that its
stability requires a delicate balance between the inter-layer and
intra-layer interactions. Similar to the 2DES case
\cite{Suen.PRL.1994}, this observation strongly favors a two-component
state.

Second, despite the apparent differences between the parameters of the
2DESs and 2DHSs in Fig. 4(a), it turns out that both systems in fact
have very similar charge distributions when the $\nu=1/2$ FQHE is
stable. In Fig. 4(a) note that the density range where we observe the
FQHE in the $W=40$ nm hole QW is similar to the range where the FQHE
is seen in 2DESs but with $W=74$ nm, almost twice wider. In Figs. 4(b)
and (c) we show the calculated charge distributions for electrons and
holes, in 74- and 40-nm-wide QWs, respectively, both at a density of
$1.50$ when the $\nu=1/2$ FQHE is strong. The calculations were done
(at $B=0$) by solving the Schroedinger and Poisson equations
self-consistently; for the 2DHS, we used the $8\times 8$ extended Kane
model \cite{Winkler.SOC.2003}. Despite the narrower well-width, thanks
to the much heavier hole effective mass, the charge distribution of
the holes is indeed bilayer-like and qualitatively very similar to the
electrons'.

Now the $\Psi_{331}$ state is theoretically expected to be stable when
the intra-layer and inter-layer Coulomb energies, $e{^2}/ 4 \pi
\epsilon l_{B}$ and $e^{2}/4 \pi \epsilon d$, respectively, are
comparable \cite{Yoshioka.PRB.1989, He.PRB.1993} ($l_B =
\sqrt{\hbar/eB}$ is the magnetic length, $d$ is the inter-layer
distance, and $\epsilon$ is the dielectric constant). For an ideal
bilayer carrier system (with zero layer thickness), the ratio $d/l_B$
accurately reflects the relative strengths of the intra-layer and
inter-layer Coulomb interactions and the $\Psi_{331}$ FQHE at
$\nu=1/2$ should be observable for $d/l_B\lesssim 2$. However, in a
system whose layer thickness, which we quantify as its
full-width-at-half-maximum ($\lambda$ in Figs. 4(b) and (c)), is
comparable to or larger than $l_B$, the short-range component of the
Coulomb interaction, which is responsible for the FQHE, softens
\cite{Shayegan.PRL.1990, He.PRB.1990}.  Associating the $\nu=1/2$ FQHE
with the $\Psi_{331}$ state, it is thus not surprising that we see the
FQHE in 2DHSs at a smaller $d/l_B\simeq 3.6$ compared to 2DESs
($d/l_B\simeq 6$) \cite{Shabani.PRB.2013}: The short-range component
of the intra-layer interaction is stronger for the 2DHS
($\lambda/l_B\simeq 1.4$) compared to the 2DES ($\lambda/l_B\simeq
2.7$); therefore to ensure the proper intra-layer to inter-layer
interaction ratio which favors the $\Psi_{331}$ state, a relatively
stronger inter-layer interaction (larger $e^2/ 4 \pi \epsilon d$) is
also needed, implying a smaller $d/l_B$ \cite{Suen.PRL.1994,
  Shabani.PRB.2013}. In a sense, thanks to its smaller layer
thickness, the 2DHS in a wide GaAs QW is closer to an ideal bilayer
system.

Revisiting the phase diagram in Fig. 4(a), the empirical shift of the
2DHS data points by 34 nm clearly leads to a close matching of the
regions where different phases (FQHE, compressible, and IP) are
observed in the 2DESs and 2DHSs \cite{Note2}. Absent of course is a
rigorous theoretical justification for this shift which results in
such a remarkable match.

\begin{figure}
\includegraphics[width=.45\textwidth]{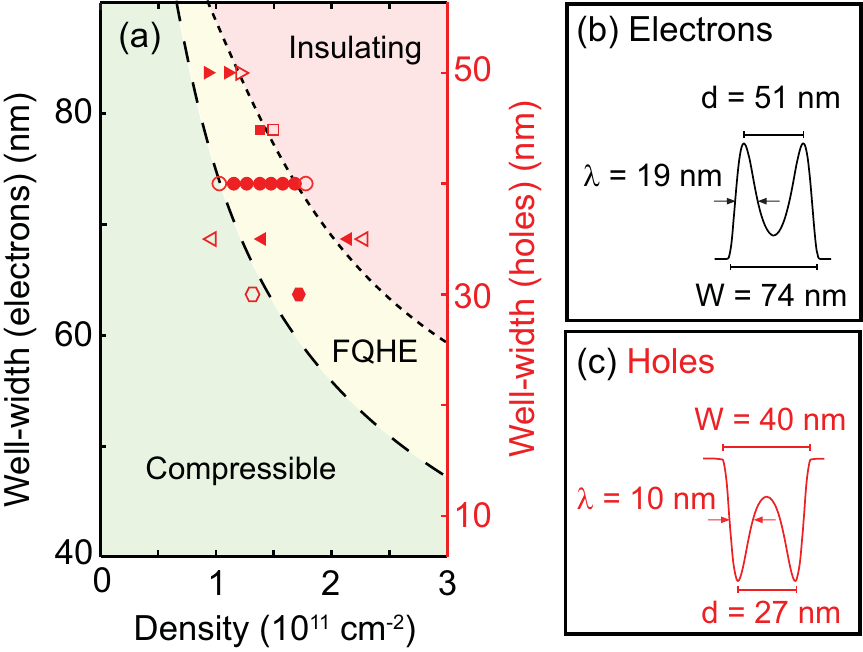}
\caption{\label{fig:waterfall2}(color online) (a) The well-width vs
  density phase diagram for different ground states observed at
  $\nu=1/2$ in 2D \emph{electron} systems confined to wide GaAs QWs
  (left axis) \cite{Shabani.PRB.2013}. The yellow, red and green
  regions are where FQHE, insulating and compressible phases are
  observed, respectively. The red symbols (right axis, which is
  shifted upward by 34 nm relative to the left axis) are data points
  of 2DHSs. The solid symbols mark the parameters at which we observe
  the $\nu=1/2$ FQHE in 2DHSs, and the open symbols are when the 2DHSs
  become compressible at low $p$ or insulating at high $p$. The right
  panels show the charge distributions calculated for: (b) electrons
  at density 1.5 confined to a 74-nm-wide
  GaAs QW, and (c) holes at the same density to a 40-nm-wide GaAs
  QW. Despite the very different well-widths, the charge distributions
  are very similar for holes and electrons. In (b) and (c), we also
  mark the values of the inter-layer distance $d$ and the "layer
  thickness" $\lambda$. }
\end{figure}

\begin{figure}
\includegraphics[width=.45\textwidth]{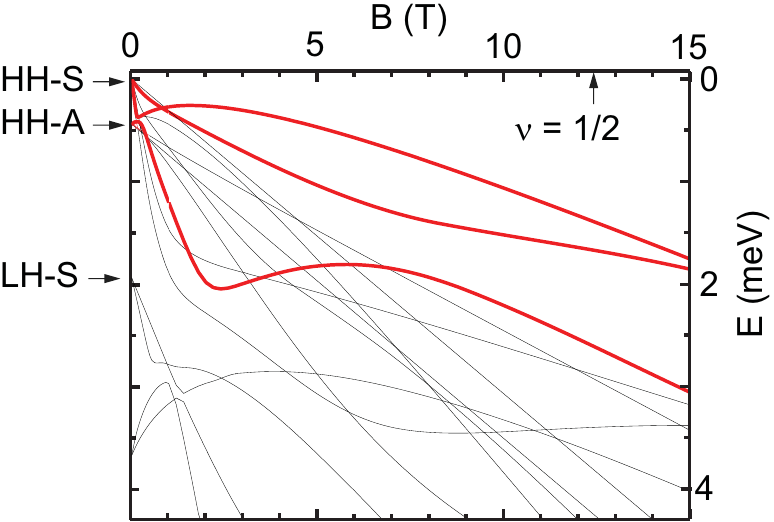}
\caption{\label{fig:waterfall2}(color online) Landau level fan diagram
  calculated for a 2DHS confined to a 40-nm-wide QW at $p = 1.5$. The
  three lowest-energy levels, which have $N=0$ character at high
  fields near $\nu=1/2$ ($B=12.3$ T), are shown in red. }
\end{figure}

Finally, we address another relevant energy scale, namely the
pseuso-spin energy splitting. In 2DESs, the pseudo-spins are the
symmetric (S) and anti-symmetric (A) electric subbands. Their energy
separation, which is essentially the "inter-layer tunneling" energy,
remains fixed as a function of $B$ in the absence of interaction. In
2DHSs the situation is more complex because of the coexistence and
mixing of HH and LH states. We illustrate this in Fig. 5 where we show
the energy vs $B$ LL diagram, calculated for a 2DHS confined to a
symmetric 40-nm-wide QW at density $p = 1.5$ \cite{Winkler.SOC.2003}.
In our wide QWs with a sufficiently large density, the lowest two
subbands are both HH-like near the subband edge. Yet $B$ mixes the LH
states into the wavefunction such that, at high field near $\nu =
1/2$, the three lowest-energy LLs (marked in red in Fig. 5) have $N=0$
character.  The pseudo-spin energy splitting is therefore of the order
of the separation between these energy levels at $B=12.3$ T which is
smaller than $\simeq 10$ K, or $\simeq 0.06$ in units of
$e^2/4\pi\epsilon l_B$. This is comparable to the values for 2DESs
when $\nu=1/2$ FQHE is seen at similar densities \cite{Suen.PRL.1994,
  Shabani.PRB.2013}, and is consistent with the $\Psi_{331}$ state
when the intra-layer Coulomb energy dominates over the pseudo-spin
energy separation.

\begin{acknowledgments}
  We acknowledge support from the DOE BES (DE-FG02-00-ER45841) for
  measurements, and the Gordon and Betty Moore Foundation (GBMF2719),
  the Keck Foundation, and the NSF (DMR-1305691 and MRSEC
  DMR-0819860) for sample fabrication. Work at Argonne was supported
  by DOE BES under Contract No. DE-AC02-06CH11357. A portion of this
  work was performed at the National High Magnetic Field Laboratory,
  which is supported by NSF Cooperative Agreement No. DMR-1157490, by
  the State of Florida, and by the DOE. We thank J. K. Jain and
  Z. Papic for illuminating discussions, and S. Hannahs, E. Palm,
  J. H. Park, T. P. Murphy, and G. E. Jones for technical assistance.
\end{acknowledgments}

\bibliography{../bib_full}
\end{document}